\documentclass[a4paper,12pt]{article}
\usepackage[cp1251]{inputenc}
\usepackage[ukrainian, russian, english]{babel}

\usepackage{graphicx}
\usepackage{epstopdf}
\usepackage{amsmath,amssymb,euscript,amsthm}
\usepackage{placeins}
\theoremstyle{plain}

\theoremstyle{definition}

\theoremstyle{remark}
\newtheorem{remk}{Remark}
\newtheorem{prop}{Proposition}

\begin{document}

\begin{center}

{\vspace{5mm}\fontsize{20pt}{12pt}\selectfont\textbf{  Dynamics of ternary statistical experiments \\
with equilibrium state}} 
\vskip12pt
\textbf{M.L.Bertotti$^\flat$, S.O. Dovgyi$^\sharp$, D. Koroliouk$^\sharp$ }\\
$^\flat$Free University of  Bozen-Bolzano, Faculty of Science and Technology,
Piazza Universita' 5, 39100 Bozen-Bolzano, Italy.\\
$^\sharp$Institute of telecommunications and global information space Ukr.Acad.Sci., Chokolovskiy Boulevard 13, 03110 Kiev, Ukraine.
\end{center}

\vskip18pt
\par\textbf{Abstract.}
We study the scenarios of the dynamics of ternary statistical experiments, modeled by means of difference equations. The important features are a balance condition and the existence of a steady-state (equilibrium).
We give a classification of scenarios of the model’s evolution which are significantly different between them, depending on the domain of the values of the model basic parameters $V_0$  and $\rho_0$ (see Proposition 1).


\section{Introduction}

Equilibrium processes are common in complex systems and play important role in many mechanisms of interaction and self-regulation. Studying these processes, one should develop an adequate method of description, analysis and prediction of the behavior of such systems, taking into account the action of a wide variety of external factors.
In the task of monitoring the dynamics of equilibrium system , one can study the frequencies (of concentrations) of a fixed set of alternative features in increasing discrete time instants (also called stages).

Such a frequency model has numerous practical interpretations such as the concentration of a substance in chemical reactions, the establishment or breaking of chemical bonds between molecules, the interaction between the elements that make up a complex system, the learning process of a team, etc.

In particular, if there are three alternatives, the model is called a ternary statistical experiment and determines the frequency of occurrence of one of the three possible attributes. So the sum of the three frequencies is always 1.

The existence of three alternative features means that the analysis can actually be carried out for two independent frequencies. However, for the sake of analysis symmetry, it is advisable, in the case of three frequencies, to analyze vector statistical experiments, which dimension coincides with the total number of features. For this reason, ternary statistical experiments with three possible attributes are defined by three-dimensional vectors.

In this setting, the mathematical model can be built as statistical experiments in which the frequency of the presence of attributes depends on the set of their frequency in the system at the previous stage.

The equilibrium of a system is determined by the form of this dependence, which is called the regression function with directing parameters that provide a change the frequency of a given attribute in proportion to its frequency simultaneously with a change in the frequencies of other attributes (alternatives).

In the works \cite{DK_6},  \cite{DK_8},  \cite{KKR},  \cite{DK_19},  \cite{DK_13} a model of binary statistical experiments with persistent regression and equilibrium is considered for several aspects of investigation. The present work considers the ternary model in a three-dimensional scheme of discrete stationary Markov diffusion, defined by a vector difference stochastic equation. The classification of ternary statistical experiments limit dynamics is given in base of specific law of large numbers by passing to a double limit by the sample volume and by the discrete time parameter.

\section{Building a model}

We consider \textit{statistical experiments} (SE) with persistent linear regression \cite{DK_6} with additional alternatives.

The basic idea of the model construction is to choose a main factor that determines the essential state of SE, supplemented by additional alternatives in the way that the aggregation of the principal factor and its complementary alternatives completely describe the dynamics of SE on time.

The \textit{basic characteristic} of the main factor and of the additional alternatives are their probabilities (frequencies): $P_0$ of the main factor and  $P_1, \ P_2$  of the additional alternatives, for which the balance condition takes place:
\begin{equation}
\label{eq1}
P_0+P_1+P_2=1.
\end{equation}

The dynamics of SE characteristics is determined by a \textit{linear regression function} \cite{DK_8} which specifies the values of SE characteristics in the next stage of observation, for given value probability at the present stage.

Consider a sequence of SE characteristics values which depends on the stage of observation, or, equivalently, on a discrete time parameter $k\geq0$:

\begin{equation}
\nonumber
P(k):=(P_0(k), P_1(k), P_2(k)) \ , \ \ k\geq0,
\end{equation}
and their increments at  $k$-th time instant:
\begin{equation}
\nonumber
\Delta P(k+1):=P(k+1)-P(k) \ , \ \ k\geq0.
\end{equation}
The linear regression function of increments is defined by a matrix which is generated by directing action parameters:
\begin{equation}
\label{eq2}
\Delta P(k+1)=-\widehat{\mathbb{V}}P(k) \ , \ \ k\geq0,
\end{equation}
where
\begin{equation}
\label{eq3}
\begin{split}
&\widehat{\mathbb{V}}:=[\widehat{V}_{mn} \ ; \ \ 0\leq m,n\leq 2],\\
&\widehat{V}_{mm}=2V_m \ , \ \ \widehat{V}_{mn}=-V_n \ , \ \ 0\leq n\leq 2 \ , \ \ n\neq m.
\end{split}
\end{equation}

The directing action parameters $V_0, V_1, V_2$ satisfy the following inequality \cite{DK_6}:
\begin{equation}
\label{eq4}
|V_m|\leq1 \ , \ \ 0\leq m\leq2.
\end{equation}
An important feature of SE is the presence of a steady state $\rho$ (equilibrium), which is determined by zero of the regression function of increments :
\begin{equation}
\label{eq5}
\widehat{\mathbb{V}}\rho=0,
\end{equation}
or in scalar form:
\begin{equation}
\label{eq6}
\widehat{V}_m\rho:=\widehat{V}_{m0}\,\rho_0+\widehat{V}_{m1}\,\rho_1+
\widehat{V}_{m2}\,\rho_2=0 \ , \ \ 0\leq m\leq2.
\end{equation}
Of course, the following balance condition takes place:
\begin{equation}
\label{eq7}
\rho_0+\rho_1+\rho_2=1.
\end{equation}
Next, we consider the fluctuations probabilities relative to equilibrium value
\begin{equation}
\label{eq8}
\widehat{P}_m(k):=P_m(k)-\rho_m \ , \ \ 0\leq m\leq2.
\end{equation}
\underline{\textbf{The basic assumption.}} The SE dynamics is determined by a difference equation for the main factor probabilities $\widehat{P}_0(k)$, and by the probabilities of additional alternatives $\widehat{P}_1(k)$ and $\widehat{P}_2(k)$
\begin{equation}
\label{eq9}
\Delta\widehat{P}(k+1)=-\widehat{\mathbb{V}}\widehat{P}(k) \ , \ \ k\geq0,
\end{equation}
or in scalar form:
\begin{equation}
\label{eq10}
\Delta\widehat{P}_m(k+1)=\widehat{V}_{m0}\,\widehat{P}_0(k)
+\widehat{V}_{m1}\,\widehat{P}_1(k)+\widehat{V}_{m2}\,\widehat{P}_2(k) \ , \ \ 0\leq m\leq2 \ , \ \ k\geq0.
\end{equation}
Also the initial values have to be fixed:
\begin{equation}
\nonumber
\widehat{P}(0)=(\widehat{P}_0(0),\widehat{P}_1(0),\widehat{P}_2(0)).
\end{equation}

\begin{remk}
\label{remk1}
Considering equations (5) - (6) and the balance condition (7), we have explicit formulas for equilibrium:
\begin{equation}
\nonumber
\begin{split}
&\rho_m=V_m^{-1}/\overline{V} \ , \ \ 0\leq m\leq2,\\
&\overline{V}:=V_0^{-1}+V_1^{-1}+V_2^{-1},
\end{split}
\end{equation}
or in other form:
\begin{equation}
\label{eq11}
\begin{split}
&\rho_0=V_1V_2/V \ , \ \ \rho_1=V_0V_2/V \ , \ \rho_2=V_0V_1/V, \\
&V:=V_1V_2+V_0V_2+V_0V_1.
\end{split}
\end{equation}
\end{remk}
The validity of the formulas (11) can be easily confirmed by their substitution in equations (6) - (7). Obviously this holds true under the additional condition: $V\not=0$.
\begin{remk}
\label{remk2}
The dynamics determination by the linear regression function (9) - (10) in regression model of statistic experiments, \underline{does not envolves} the balance condition (1), and the equilibrium (5) with additional restrictions:
\begin{equation}
\nonumber
0\leq P_m(k)\leq1 \ , \ \ 0\leq m\leq2 \ , \ \ k\geq0 \ ; \ \ 0\leq\rho_m\leq1 \ , \ \ 0\leq m\leq2
\end{equation}
for solutions of difference equations (2) or, equivalently (9) and equations (5) - (7) for equilibriums.
\end{remk}

\section{The model interpretation}

The model of SE is constructed in several stages. First, the main factor should be chosen, characterized by probability (or frequency, concentration etc.). So there exist supplementary alternatives, whose probabilities are complement to the main factor probability. In particular, having only one alternative, the classification of binary models are considered in \cite{DK_14} (see also \cite{Ara}, \cite{ScorHop}).
The presence of two or more alternatives brings more difficulties in the analysis of SE.

With a full set of SE characteristics, the probabilities of the main factor and of additional alternatives satisfy the balance condition (1) or, equivalently, the balance condition (7) and the dynamics of the main factor probability ${P}_0$, as well as of supplementary factors ${P}_1$, ${P}_2$ is given by the following difference equations for the probabilities of fluctuations for all $k\geq0$:
\begin{equation}
\begin{split}
\label{eq12}
&\Delta\widehat{P}_0(k+1)=V_1\widehat{P}_1(k) +V_2\widehat{P}_2(k)-2V_0\widehat{P}_0(k) \ ,\\
&\Delta\widehat{P}_1(k+1)=V_0\widehat{P}_0(k) +V_2\widehat{P}_2(k)-2V_1\widehat{P}_1(k) \ ,\\
&\Delta\widehat{P}_2(k+1)=V_0\widehat{P}_0(k) +V_1\widehat{P}_1(k)-2V_2\widehat{P}_2(k) \ .
\end{split}
\end{equation}
The increment of probabilities fluctuations of the main and supplementary factors
\begin{equation}
\nonumber
\Delta\widehat{P}_m(k+1):=\widehat{P}_m(k+1)-\widehat{P}_m(k) \ , \ \ 0\leq m\leq2 \ , \ \ k\geq0,
\end{equation}
is determined by the values of directing action parameters $V_0,V_1,V_2$.

\begin{remk}
\label{remk3}
The fluctuations of probabilities in (8) satisfy the balance condition:
\begin{equation}
\label{eq13}
\widehat{P}_0(k)+\widehat{P}_1(k)+\widehat{P}_2(k)=0 \ , \ \ k\geq0,
\end{equation}
and by formula (8) one has:
\begin{equation}
\label{eq14}
\Delta\widehat{P}_m(k)=\Delta{P}_m(k) \ , \ \ 0\leq m\leq2 \ , \ \ k\geq0.
\end{equation}
\end{remk}

The equation (12) characterizes two basic principles of alternatives interaction: \textit{stimulation} (positive term) and \textit{containment} (negative term).

\section{The model analysis}

The existence of an equilibrium point for the fluctuations increments regression function (5) provides the possibility to analyze the dynamics of SE (by $k\to\infty$) in view of the possible directing parameter values which satisfy the constraint (4).

The dynamics of the main factor probability is described by \textit{several scenarios}.

\begin{prop}
\label{prop1}
The main factor probability  $P_0(k)$, $k\geq0$, determined by the solution of the difference equation  (12), as well as by the basic assumption (9), with equilibrium  (11), changes when $k\to\infty$ according to the following scenarios:

\underline{\textbf{Attractive equilibrium}}: $V_0>0, \ 0<\rho_0<1$:
\begin{equation}
\label{eq15}
\lim_{k\to\infty}P_0(k)=\rho_0;
\end{equation}

\underline{\textbf{Repulsive equilibrium}}: $V_0<0, \ 0<\rho_0<1$:
\begin{equation}
\label{eq16}
\lim_{k\to\infty}P_0(k)=\begin{cases} 1 & \text{if}  \ \ \ P_0(0)>\rho_0; \\
0 & \text{if}  \ \ \ P_0(0)<\rho_0. \end{cases}
\end{equation}

\underline{\textbf{Dominant equilibrium}}:   $\rho_0 \not\in(0,1), \ V_0<0$:
\begin{equation}
\label{eq17}
\lim_{k\to\infty}P_0(k)=1;
\end{equation}

\underline{\textbf{Degenerate equilibrium}}: $\rho_0 \not\in(0,1), \ V_0>0$:
\begin{equation}
\label{eq18}
\lim_{k\to\infty}P_0(k)=0;
\end{equation}
\end{prop}

\begin{remk}
\label{remk4} Of course, the main factor dynamics scenarios can be formulated by domain of values of the directing parameters $V_0, V_1, V_2$.
\end{remk}

\begin{remk}
\label{remk5}
Similar scenarios for additional alternative dynamics take place by considering the values of parameters   $V_1,\rho_1$ or $V_2,\rho_2$.
\end{remk}

\section{Conclusion}

A model of ternary statistical experiments is introduced, which equilibrium is determined in terms of the probability dynamics of the main factor. The present work considers the ternary model in a scheme of discrete stationary Markov diffusion, defined by a difference stochastic equation. The classification of ternary statistical experiments limit dynamics is given in base of specific law of large numbers by passing to a double limit by the sample volume and by the discrete time parameter.

The classification is briefly summarized in the following figures.

\begin{figure}[h!]
\center{\includegraphics[width=0.7\linewidth]{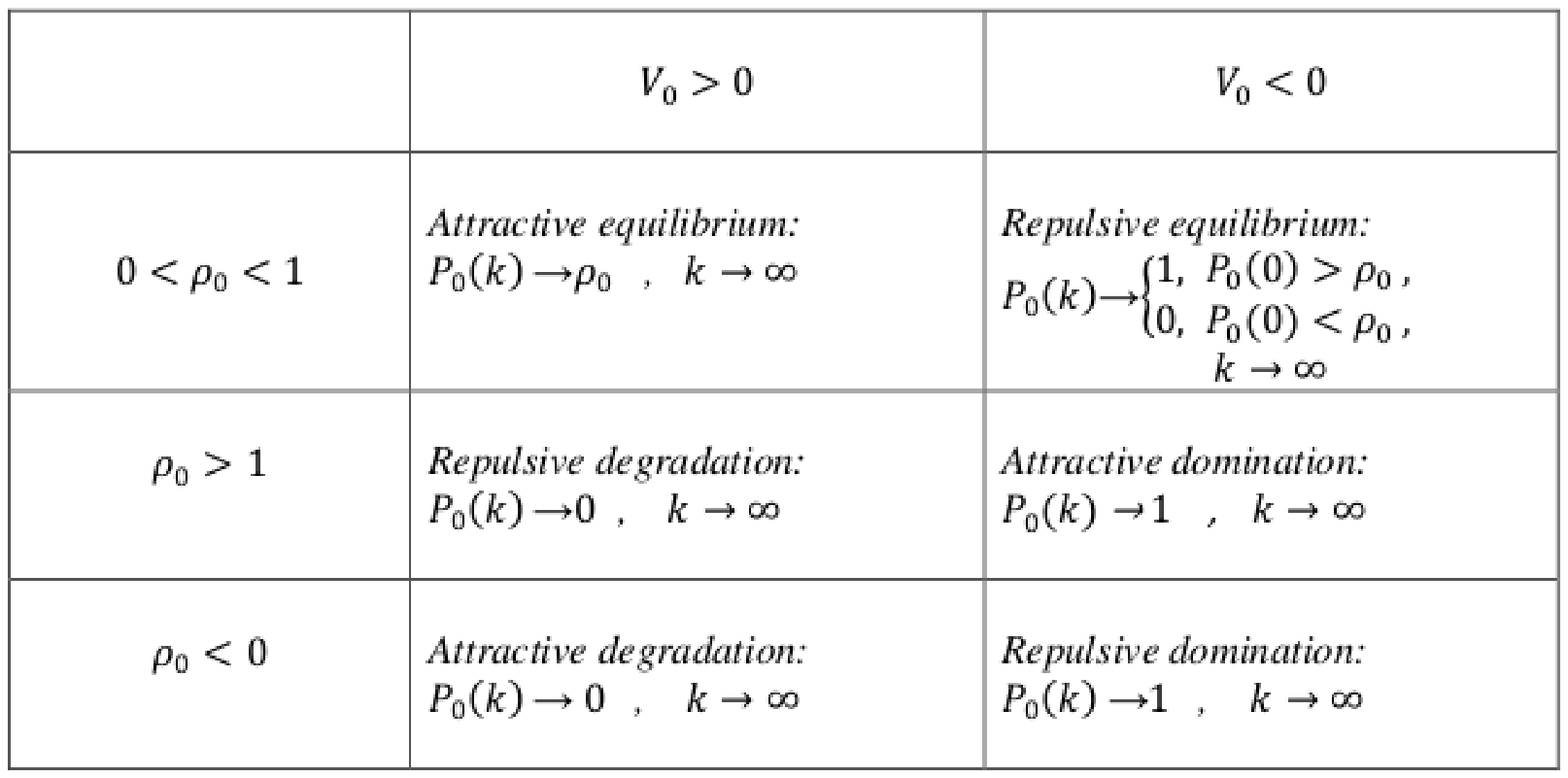}}
\caption{ Table of scenarios}
\label{Fig:1}
\end{figure}

\begin{figure}[h!]
\center{\includegraphics[width=0.7\linewidth]{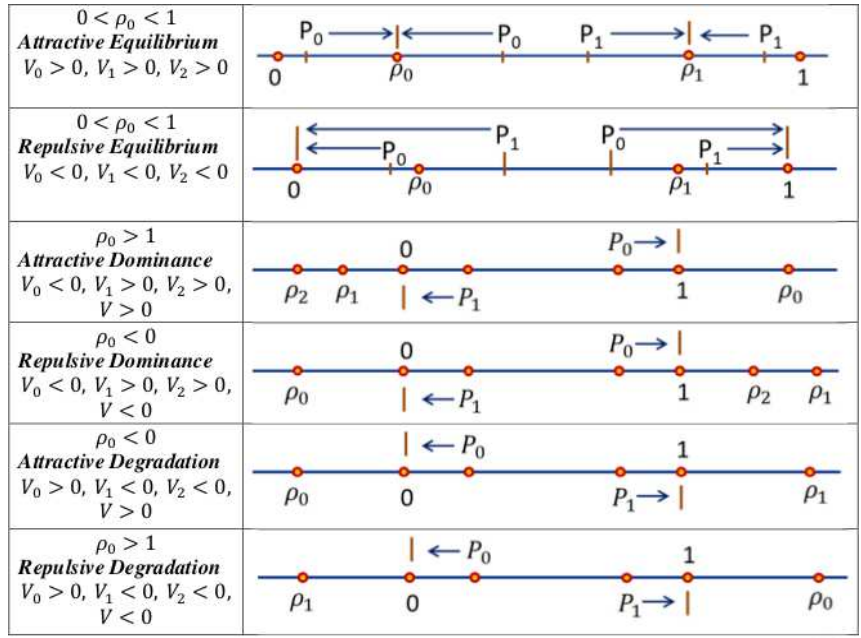}}
\caption{ Illustration of $P_i$ limit behaviour}
\label{Fig:2}
\end{figure}

\end{document}